\documentclass[preprint,prl,aps,amsmath,amssymb,color,superscriptaddress]{revtex4}
\usepackage{stmaryrd}
\usepackage{amsmath}
\usepackage{amssymb}
\usepackage{graphicx}
\usepackage{dcolumn}
\usepackage{bm}
\usepackage{tabularx}
\usepackage{color}
\begin{document}
\begin{titlepage}
\title{First-principles perspective on full-spectrum infrared photodetectors from doping an excitonic insulator}
\author{Jing Liu}
\affiliation{Key Lab of advanced optoelectronic quantum architecture and measurement (MOE), and Advanced Research Institute of Multidisciplinary Science, Beijing Institute of Technology, Beijing 100081, China}
\author{Yuanchang Li}
\email{yuancli@bit.edu.cn}
\affiliation{Key Lab of advanced optoelectronic quantum architecture and measurement (MOE), and Advanced Research Institute of Multidisciplinary Science, Beijing Institute of Technology, Beijing 100081, China}
\date{\today}

\begin{abstract}
Innovations in imaging technology involves finding strategies and materials suitable for detection applications over the entire infrared range. Herein, we propose a new design concept based on the unique feature of an excitonic insulator, namely, negative exciton transition energy ($E_t$). We demonstrate this concept using first-principles $GW$-BSE calculations on one-dimensional organometallic wire (CrBz)$_\infty$. The pristine (CrBz)$_\infty$ exhibits an excitonic instability due to a negative $E_t$ for the lowest exciton. Substitutional doping can continuously tune the $E_t$ from $\sim$0 to $\sim$0.6 eV, which shows the ability of photon detection from terahertz to near-infrared. This type of detectors have advantages of outstanding wavelength selectivity, reduced thermal disturbance and elevated working temperature. Our work not only adds another member in the family of rare one-dimensional excitonic insulators, but also opens a new avenue for the development of high-performance infrared photodetectors in the future.
\end{abstract}

\maketitle
\draft
\vspace{2mm}
\end{titlepage}

Infrared photon detection is of utmost importance for a wide range of applications, including night vision, bioimaging, communications, remote sensing, environmental monitoring, security checking, radio astronomy, etc\cite{Rogalski,LiuJ,Guo,Downs,Martyniuk,Smith,Norton}. Present-day photodetectors rely primarily on semiconductors' bandgap which determines the minimum photon energy to convert into electrical signals for subsequent processing. As a result, narrow gap systems are chosen to fabricate high-performance infrared detectors. Hg$_{1-x}$Cd$_x$Te alloy is nowadays the leading compound for applications over almost the entire infrared range from short (1 $-$ 3 $\mu$m) to very long (14 $-$ 30 $\mu$m) wavelength (corresponding to the photon energy of 0.04 $-$ 1.24 eV), ascribed to its continuously adjustable bandgap from negative of HgTe to positive of CdTe with the increase of $x$\cite{Martyniuk,Rogalski}. Nevertheless, Hg$_{1-x}$Cd$_x$Te is restrained by some fundamental limitations including high manufacturing costs, complicated synthetic process, cryogenic operating environment and precise composition control, in addition to material toxicity\cite{Rogalski,LiuJ,Smith,Norton}. There has been an ever-increasing demand for finding new design concepts and/or alternative versatile materials that can compete with and eventually replace Hg$_{1-x}$Cd$_x$Te technologies. For this purpose, a photoresponse in the whole infrared band is a prerequisite. Emerging gapless materials such as graphene and topological-semimetals are thus demonstrated for ultra-broadband photon detection, but the performance is substantially hampered by low overall quantum efficiency and high dark current\cite{LiuJ,Xia}.

In reality, the material's optical response must involve excitons (bound electron-hole pairs by Coulomb interaction), because excitons are first created from light harvesting and their subsequent charge separation produces current. The excitons are often ignored due to their small binding energy in traditional narrow-gap semiconductors, in which the excitons, once formed, would readily dissociate via thermal activation\cite{Iotti,Zrenner}. Instead, if the exciton effect is sufficiently significant, photodetectors can be fully designed based on the exciton\cite{Lukman}.

Excitonic insulator, proposed some half century ago, has a many-body ground state characterized by spontaneously formed exciton condensate\cite{Kohn}. It is viewed as an analog to the BCS superconductor and expected to host amazing macroscopic quantum properties like super transport\cite{Eisenstein,Kozlov,Sun}. Although, to date, there is no universally accepted excitonic insulator, recent advances in low-dimensional materials have opened new avenues for the search and discovery of such insulators\cite{Du,usEI,usgraphone,VarsanoNC,usHEI,usDong,VarsanoNT,Sethi}. The excitonic insulator emerges when the exciton binding energy ($E_b$) exceeds the one-electron bandgap ($E_g$), so the lowest characteristic exciton has an inherently negative transition energy ($E_t = E_g - E_b$) \cite{Kohn}. With this unique property, the excitonic insulators offer an unprecedented opportunity for exciton-based full-spectrum infrared detection because they hold the potential to allow continuous regulation of $E_t$ from negative to positive, just as continuous regulation of $E_g$ in Hg$_{1-x}$Cd$_x$Te alloys. This is unusual as a gapped system generally presents a positive $E_t$, which defines the longest wavelength of detection. In this light, the excitonic insulators are among the most attractive alternatives to replace Hg$_{1-x}$Cd$_x$Te for the infrared detection. It has to be noted, however, that the characteristic exciton in an excitonic insulator is optically inactive\cite{Halperin,usEI,usPRL}, and therefore, stimulating its optical activity represents another necessary prerequisite for the realization of excitonic-insulator infrared detection.

In this work, through first-principles $GW$ plus Bethe-Salpeter equation (BSE) calculations, we illustrate the strategy of designing exciton-based full-spectrum infrared detector by substitutionally doping an excitonic insulator. First, we show the occurrence of excitonic instability in one-dimensional organometallic sandwich chromium-benzene wire (CrBz)$_\infty$, whose characteristic exciton has a negative $E_t$ of -0.27 eV and a giant $E_b$ up to 1.08 eV. Then we investigate the effect of substitutional doping (H$\rightarrow$F and C$\rightarrow$N) on the characteristic exciton. We find a gradual blue-shift of $E_t$ from negative to around zero, and to positive, depending on the dopant concentration and distribution, while maintaining $E_b$ $>$ 0.5 eV. Moreover, the doping causes symmetry-breaking and accordingly, the characteristic exciton is optically activated. All these together indicate the feasibility of building an exciton-based full-spectrum infrared detector that can work at high temperature. Finally, we discuss the advantages of here studied exciton-based detectors over traditional bandgap-based ones in terms of colour selectivity and reduced thermal disturbance.

All DFT calculations were performed using the Vienna \emph{ab initio} simulation package (VASP)\cite{vasp} within the Perdew-Burke-Ernzerhof (PBE) exchange correlation functional\cite{PBE}. Projector augmented wave (PAW)\cite{PAW} method was used with a cutoff kinetic energy of 500 eV. A supercell model was adopted for the isolated (CrBz)$_\infty$ and the separation between two neighboring wires was at least 12 \AA. Single-shot $G_0W_0$ calculations were carried out for quasi-particle band structure\cite{Hybertsen} with a \emph{k}-mesh density no less than 14/\AA\ and a total of more than 6 times the number of valence bands. The BSE is solved for excitonic properties\cite{Rohlfing} on top of the $G_0W_0$ results, with three valence and four conduction bands included for the BSE Hamiltonian. Exciton wave-functions were obtained from the Yambo code\cite{yambo}.

\begin{figure}[tbp]
	\includegraphics[width=0.85\columnwidth]{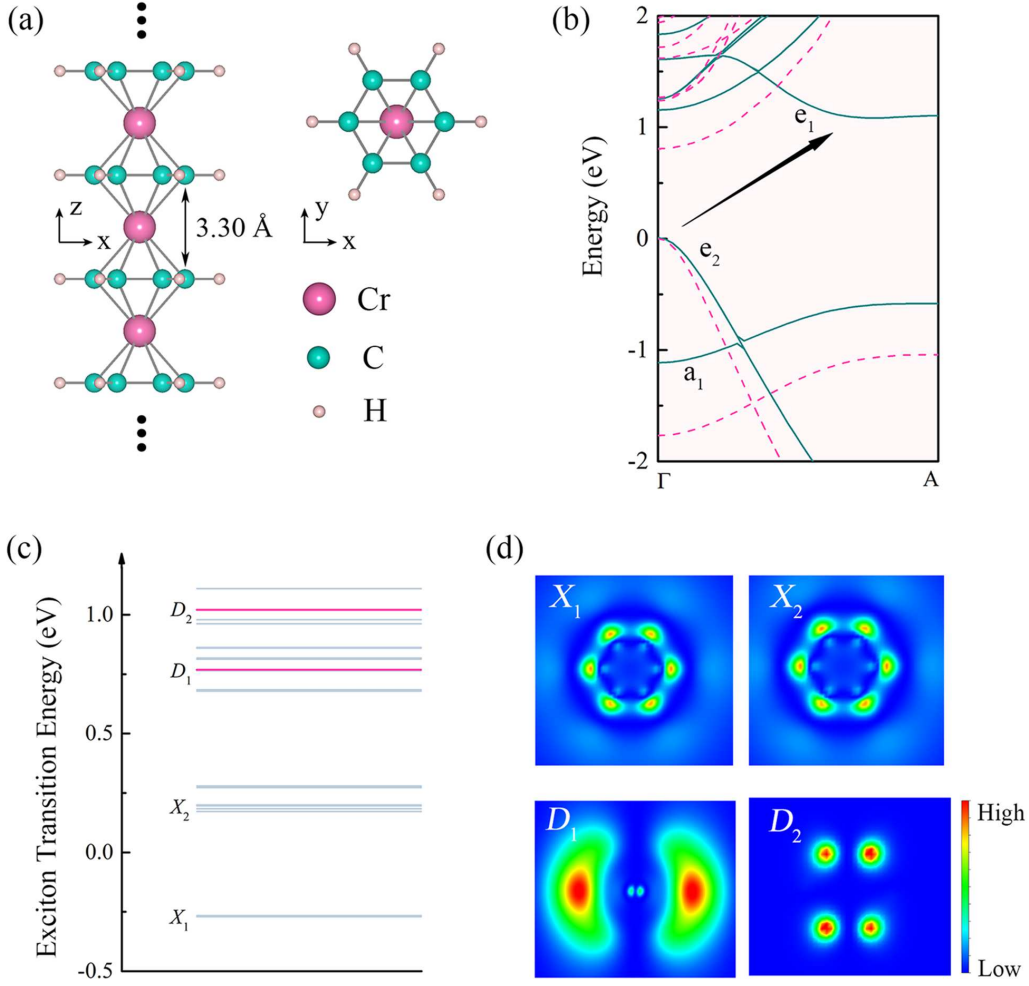}
	\caption{\label{Fig1} (Color online) (a) Crystal structure, (b) PBE (cyan solid lines) and $G_0W_0$ (pink dashed lines) bands, and (c) transition energies of low-energy excitons for the (CrBz)$_\infty$ wire. In (b), the valence band maximum is set as energy zero and the arrow denotes the minimum gap by the PBE. In (c), $D$- and $X$- series represent the bright (pink lines) and dark (gray lines) excitons, respectively. (d) Cross section of the real-space exciton wave-functions in the $xy$ plane for the states marked in (c).}
\end{figure}

One-dimensional (CrBz)$_\infty$ wire consists of alternately arranged Cr atoms and benzene (Bz) rings, as plotted in Fig. 1(a). Its ground-state structure possesses the $D_{6h}$ symmetry\cite{Xiang}. Under such a crystal field, the Cr 3$d$-orbitals split into $e_1$ ($d_{xz}$ and $d_{yz}$ orbitals) and $e_2$ ($d_{x^2-y^2}$ and $d_{xy}$ orbitals) doublets, and $a_1$ ($d_{z^2}$ orbital) singlet\cite{LiJCP,LiJPCC}. Due to the Cr 3$d^4$4$s^2$ electronic configuration, the lower-lying $a_1$ singlet and $e_2$ doublet are fully occupied while the $e_1$ doublet is unoccupied. So (CrBz)$_\infty$ is expected to be a non-spin-polarized semiconductor with an $e_1$ and $e_2$ doublet separated gap. Our PBE calculations verify this picture and give an indirect gap of 1.08 eV, as shown by the arrow in Fig. 1(b). But this is contrast to previous report of a direct gap 0.71 eV at $\Gamma$ using the PW91 functional\cite{Xiang}. Our high-level theoretical calculations by the $GW$ yields a direct gap of 0.81 eV at $\Gamma$, which is characterized by Cr $e_2$ doublet and Bz $\pi$-orbital [see pink dashed lines in Fig. 1(b)]. Indirect to direct gap transition from the PBE to the $GW$ is probably attributed to disproportionate quasi-particle corrections to the localized $d$- and delocalized $\pi$-orbitals\cite{usMnCp}.

We then solve the BSE for the excitonic properties and the $E_t$ of low-energy excitons are summarized in Fig. 1(c). Herein both bright and dark excitons are considered (respectively denoted as $D$- and $X$-series hereafter). The most interesting finding is a negative $E_t$ of -0.27 eV for the lowest $X_1$ exciton, which means its spontaneous formation and the occurrence of excitonic instability. Therefore, the (CrBz)$_\infty$ can serve as a rare candidate for an intrinsic one-dimensional excitonic insulator. To our best knowledge, carbon nanotubes are the only case proposed to host such an amazing one-dimensional excitonic insulating state\cite{VarsanoNC}.

In practice, only the bright excitons have optical response and are useful for light absorption and detection while the dark excitons are not. However, the characteristic exciton that leads to instability in excitonic insulators is generally dark\cite{Halperin,usEI}, like here the $X_1$. This apparent contradiction has to be solved before realizing the photon detection based on the excitonic insulators. According to the knowledge of group-theory, optical inactivity of the $X_1$ is resulted from the dipole forbidden transition between band-edge states in terms of the unique $D_{6h}$ symmetry of the (CrBz)$_\infty$. To illustrate this point more clearly, we representatively select two excitons from the $D$- and $X$-series respectively and compare their characteristics in Fig. 1(d). It is obviously seen that the $X$-series excitons all respect the $D_{6h}$ symmetry while the $D$-series excitons show different degrees of symmetry breaking.

A full-spectrum exciton-based infrared detector involves three key elements of the characteristic exciton: continuously adjustable $E_t$ almost from zero, large enough $E_b$ and optical activity. Herein we focus on the means of substitutional doping, including the replacement of C(H) with N(F), for the following four considerations: 1) Such reactions have been chemically established and result in small structural changes\cite{Adler,ussmall}, 2) It is able to modulate the materials' electronic and optical properties effectively\cite{LuJACS}, 3) It can yield symmetry-breaking, hence causing dark-to-bright exciton conversion, and 4) It affects $E_b$ relatively marginally\cite{usDong}.

\begin{table*}
	\caption{The $G_0W_0$ gap $E_g$ (eV), the exciton transition energy $E_t$ (eV) and the exciton binding energy $E_b$ (eV), corresponding to the different crystal structures inserted in Fig. 2.}
	\label{tab:group}
	\begin{ruledtabular}
		\begingroup
		\setlength{\tabcolsep}{4.5pt}
		\renewcommand{\arraystretch}{1.5}
		\begin{tabular}{cccccccc}
			
            &CrC$_5$NH$_5$  & (CrC$_5$NH$_5$)$^d_2$ & (CrC$_5$NH$_5$)$^s_2$ & Cr$_2$C$_{11}$NH$_{11}$ & Cr$_2$C$_{11}$NH$_{10}$F & CrC$_6$H$_4$F$_2$ & CrC$_6$H$_3$F$_3$ \\			
            \hline
			
			$E_g$ & 0.945 & 1.010 & 0.995 & 0.832 & 1.108 & 1.497 & 1.780 \\
			
			$E_t$ & 0.060 & 0.123 & 0.207 & 0.326 & 0.417 & 0.529 & 0.627 \\
			
			$E_b$ & 0.885 & 0.887 & 0.789 & 0.506 & 0.691 & 0.968 & 1.153 \\
		\end{tabular}
	    \endgroup
	\end{ruledtabular}
\end{table*}

\begin{figure*}[tbp]
	\includegraphics[width=0.85\columnwidth]{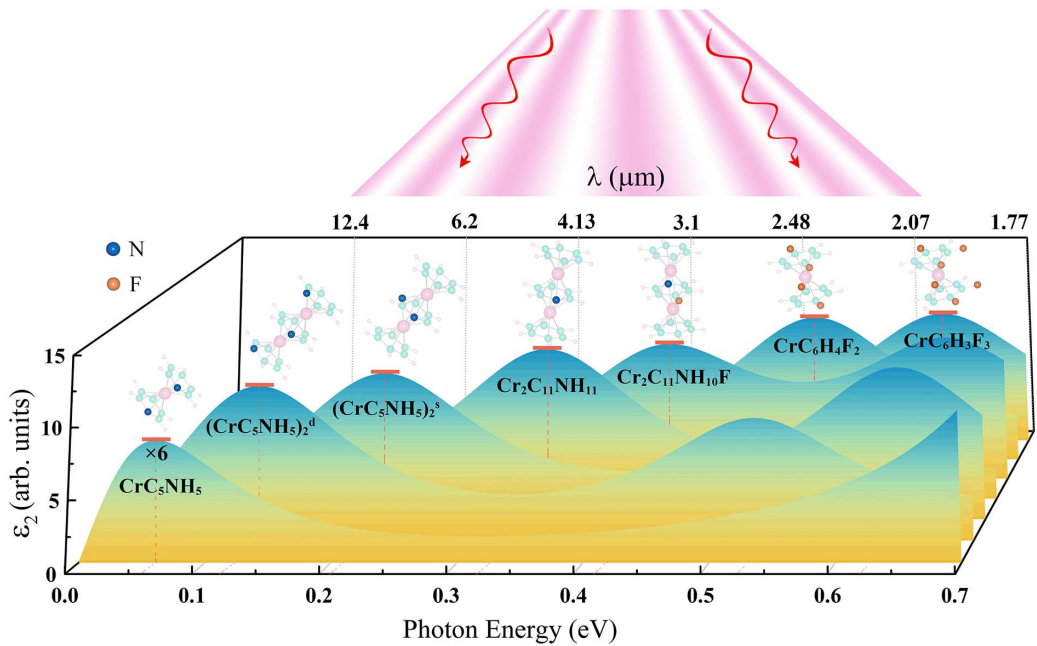}
	\caption{\label{Fig2} (Color online) Calculated imaginary part of the BSE dielectric function for different doping systems with the incident light oriented along the wire. See Table I for more details. For clarity, the intensity of CrC$_5$NH$_5$ is multiplied by six. Inserts show the crystal structures of doped (CrBz)$_\infty$.}
\end{figure*}

The calculated results of several representative substitutions are presented in Table I and Fig. 2. Indeed, the doping tunes $E_t$ of the characteristic exciton, ranging from $\sim$0 eV to $\sim$0.6 eV in terms of different dopants, concentrations and distributions, while keeping the $E_b$ $>$ 0.5 eV (see Table I). Clear absorption peak related to the characteristic exciton has also been observed from the imaginary part of the BSE dielectric function (see Fig. 2). Combination of these three central points indicates the promising application potential for full-spectrum infrared detection, which can even extend to the terahertz band and work at room temperature, as well.

Specifically, one N replacing one C gives an $E_t$ of 0.060 eV, very close to zero, which illustrates the possibility for the very far-infrared ($\sim$20 $\mu$m). Maintaining N concentration, the $E_t$ could increase to 0.123 or 0.207 eV depending on whether the two N atoms are in different or the same Bz. Halving N concentration further increases the $E_t$ to 0.326 eV. On the other hand, F replacing H tends to cause blue shift of both $E_t$ and $E_g$ (see Table I). It is worth noting that, shown here are only a few concrete examples, but in principle, any $E_t$ between zero and electronic bandgap can be obtained via appropriate doping. This fact supports our statement of an exciton-based full-spectrum photon detection.

\begin{figure}[tbp]
	\includegraphics[width=0.85\columnwidth]{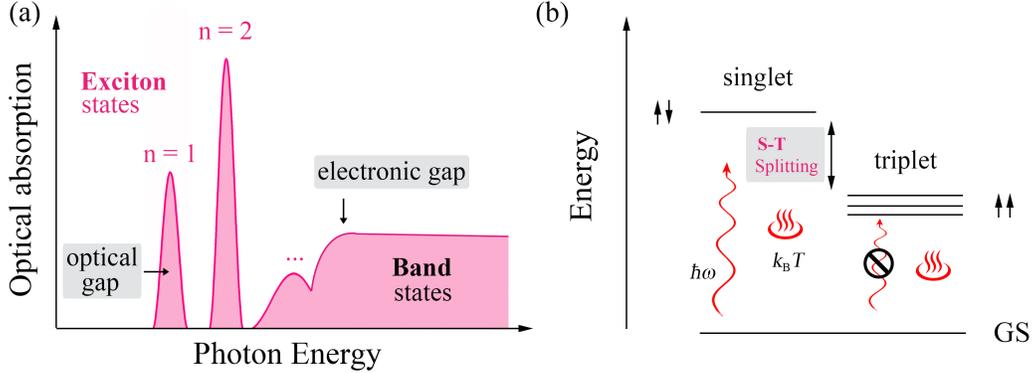}
	\caption{\label{Fig3} (Color online) (a) Typical adsorption spectrum for a low-dimensional system, in which the discrete exciton states lie below the electronic bandgap and the continuous band states lie above. Energy difference between the electronic bandgap and optical gap defines the $E_b$ of the lowest characteristic exciton. (b) Schematic illustration of the singlet-triplet splitting in terms of exciton spin, and their different optical ($\hbar\omega$) and thermal ($k_BT$) excitations.}
\end{figure}

The exciton-based infrared detectors have two prominent advantages over traditional bandgap-based ones. One is outstanding wavelength selectivity. As illustrated in Fig. 3(a), exciton absorption refers to discrete energy levels below the electronic bandgap and is limited to a very narrow band range. It is thus possible to design a photodetector that only responds to a specific wavelength. This becomes even more practical when assembling with low-dimensional materials, where the lowest exciton can be hundreds of millielectron volts away from electronic bandgap\cite{usPRL}. In sharp contrast, traditional photodetectors respond to any wavelength light in energy larger than the bandgap.

The other is reduction of the thermal disturbance. Current signal of photodetectors is produced through the charge separation of photo-excited electron-hole pairs under a built-in electric field\cite{matter}. Actually, the electron-hole pairs may also be thermally triggered, leading to an electrical output in the absence of light (dark current) and reducing the detection sensitivity. As the detection moving to very far-infrared region, thermal noise would become more pronounced and might confuse the recognition. Regarding to the bandgap-based paradigm, the longer the target wavelength is, the narrower the bandgap is required and the larger the corresponding dark current will be. For example, this poses an almost insurmountable constraint for gapless semimetals like graphene and less efficient unbiased mode has to be invoked for the charge separation\cite{LiuJ,Nathan}.

By contrast, the exciton-based paradigm no longer necessitates a narrow bandgap, which would allow the thermal excitation only relevant to occupation of the exciton state rather than the band state. Hence, the dark current could be considerably reduced in exciton-based detectors as compared to their bandgap-based counterparts. In addition, as illustrated in Fig. 3(b), considering spin degree of freedom, the excitons split into a lower-lying triplet and a higher-lying singlet with a small splitting energy. Photo-excitation is only effective for the singlet as a result of the spin-selection rule\cite{usgraphone} while thermal excitation is effective for the both. Such spin-dependent behavior would still provide us with the possibility to extract weak light signals by resolving spin information even if the thermal fluctuation noise overwhelms the photocurrent.

Next we briefly discuss two concerns regarding the practical feasibility. One is related to the fabrication of one-dimensional linear sandwich wires. As far as we know, the (CrBz)$_\infty$ has not been synthesized and no multidecker Cr$_n$Bz$_{n+1}$ clusters other than CrBz$_2$ has been produced\cite{Kurikawa}. Nevertheless, recent advances in organometallic chemistry have allowed the growth of several this kind of molecular wires with total metal atoms ranging from a few to a thousand\cite{Huttmann,Miyajima,Santhini,Nagao}, and more importantly, the developing method is believed to work for variation of either the ligand (e.g., benzene) or the metal centers (element of the 3$d$ or 4$f$ series)\cite{Huttmann}.
In this context, the preparation of (CrBz)$_\infty$ does not seem impossible and we expect our work to stimulate experimental interests.

\begin{figure}[tbh]
	\includegraphics[width=0.85\columnwidth]{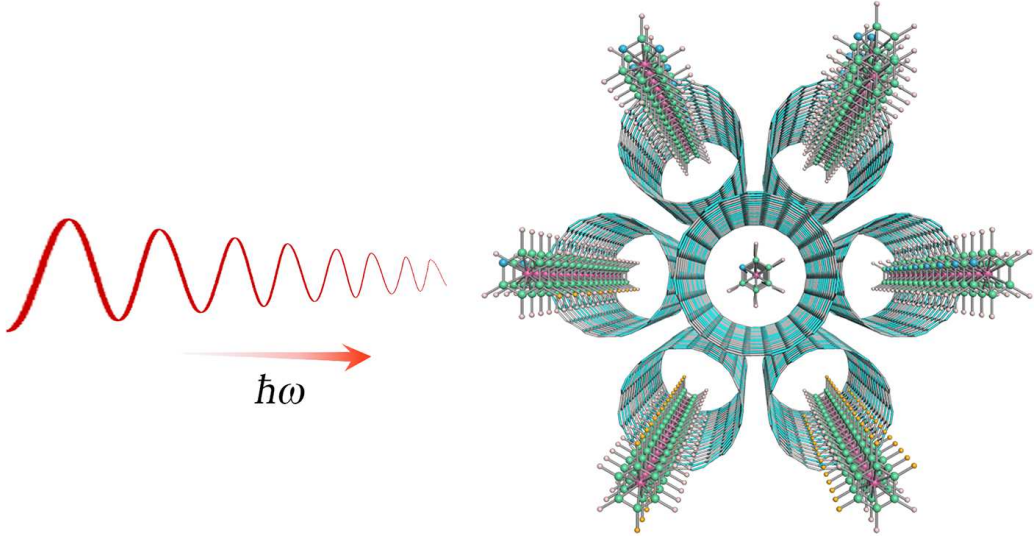}
	\caption{\label{Fig4} (Color online) A device design scheme for full-spectrum infrared detectors based on the array composed of boron nitride nanotube encapsulated doped-(CrBz)$_\infty$. Each doped-(CrBz)$_\infty$ works as an independent elementary component to detect a specific wavelength. Boron nitride nanotubes are used to reduce the influence on the exciton.}
\end{figure}

The other concern is related to the environment effect in practical applications, which is of particular importance for the excitonic properties of a one-dimensional system. To this end, we design a working scheme as schematically shown in Fig. 4. Each doped-(CrBz)$_\infty$ is encapsulated in a boron nitride nanotube to form an elementary device that only responds to a specific infrared wavelength. Then a large number of such components for different wavelengths are assembled to constitute an infrared detection array. In this way, both detection bandwidth and wavelength selectivity can be achieved. Using boron nitride nanotubes is mainly because of its large bandgap ($\sim$5 eV), low dielectric-constant and free of dangling bonds or surface charge traps\cite{usgraphone,Dean}. Not only does this help to eliminate the interwire coupling, hence maintaining the wavelength selectivity of each elementary device, but the essence of low dielectric-constant also helps to reduce the effect on excitons\cite{usgraphone}.

In conclusion, we propose a new design concept for full-spectrum infrared detection by manipulating the characteristic exciton of an excitonic insulator, and its feasibility is verified by the first principles $GW$-BSE calculations on the prototype (CrBz)$_\infty$. Unlike the traditional infrared detector, the exciton-based paradigm works by the resonant light absorption of discrete exciton states, and therefore is much superior in wavelength selectivity. In addition, it can also possess the advantages of high operating temperature and reduced thermal disturbance, going beyond classical bandgap-based one. Our work not only adds another member in the family of rare one-dimensional excitonic insulators, but also opens a new avenue for the development of high-performance infrared photodetectors in the future.

\begin{acknowledgments}
This work was supported by the Ministry of Science and Technology of China (Grant No. 2020YFA0308800) and the National Natural Science Foundation of China (Grant No. 12074034).
\end{acknowledgments}

\end{document}